\documentclass{article}

     \usepackage[nonatbib]{tccml_neurips_2020}

\usepackage[utf8]{inputenc} 
\usepackage[T1]{fontenc}    
\usepackage{hyperref}       
\usepackage{url}            
\usepackage{booktabs}       
\usepackage{amsfonts}       
\usepackage{nicefrac}       
\usepackage{microtype}      

\usepackage{graphicx}      

\title{Using attention to model long-term dependencies in occupancy behavior}

\author{
  Max Kleinebrahm \\
  Chair of Energy Economics\\
  Karlsruhe Institute of Technology\\
  Karlsruhe, Germany \\
  \texttt{max.kleinebrahm@kit.edu} \\

  \And
  Jacopo Torriti \\
  School of the Built Environment \\
  University of Reading \\
  Reading, United Kingdom \\
  \texttt{j.torriti@reading.ac.uk} \\
  
  \And
  Russell McKenna \\
  Chair in Energy Transition \\
  University of Aberdeen \\
  Aberdeen, United Kingdom \\
  \texttt{russell.mckenna@abdn.ac.uk} \\
  
  \And
  Armin Ardone \\
  Chair of Energy Economics \\
  Karlsruhe Institute of Technology\\
  Karlsruhe, Germany \\
  \texttt{armin.ardone@kit.edu} \\
  
  \And
  Wolf Fichtner \\
  Chair of Energy Economics \\
  Karlsruhe Institute of Technology\\
  Karlsruhe, Germany \\
  \texttt{wolf.fichtner@kit.edu} \\
}

\begin{document}

\maketitle

\begin{abstract}
  Models simulating household energy demand based on different occupant and household types and their behavioral patterns have received increasing attention over the last years due the need to better understand fundamental characteristics that shape the demand side. Most of the models described in the literature are based on Time Use Survey data and Markov chains. Due to the nature of the underlying data and the Markov property, it is not sufficiently possible to consider day to day dependencies in occupant behavior. An accurate mapping of day to day dependencies is of increasing importance for accurately reproducing mobility patterns and therefore for assessing the charging flexibility of electric vehicles. This study bridges the gap between energy related activity modelling and novel machine learning approaches with the objective to better incorporate findings from the field of social practice theory in the simulation of occupancy behavior. Weekly mobility data are merged with daily time use survey data by using attention based models. In a first step an autoregressive model is presented, which generates synthetic weekly mobility schedules of individual occupants and thereby captures day to day dependencies in mobility behavior. In a second step, an imputation model is presented, which enriches the weekly mobility schedules with detailed information about energy relevant \textit{at home} activities. The weekly activity profiles build the basis for modelling consistent electricity, heat and mobility demand profiles of households. Furthermore, the approach presented forms the basis for providing data on socio-demographically differentiated occupant behavior to the general public.
  
\end{abstract}

\section{Introduction}
\label{introduction}

Occupant behavior has been identified as having a significant impact on household energy consumption \cite{Steemers.2009}. Therefore, there has been an increasing research interest in the field of behavioral modelling over the last years with the aim to explain dynamics in residential energy demand based on energy related activities \cite{Torriti.2014, Torriti.2017}. A large number of studies focus on the modelling of activity sequences of single households or individuals with the objective to describe occupant behavior on an aggregated level for socio-demographic differentiated groups \cite{Richardson.2008, Wilke.2013, Flett.2016, Aerts.2014}. Time use data (TUD) are used as a data basis, which provide information on the temporal course of occupant activities over single days and are available for various countries in the form of population representative samples \cite{Eurostat.2000}. Based on occupant behavior, different approaches were developed that connect occupant activities with electrical household appliances and thus generate synthetic electricity demand profiles \cite{Yamaguchi.2018}. The aim of these studies is to gain a deeper understanding of household electricity demand in order to e.g. be able to evaluate device-specific efficiency measures, time-dependent electricity tariffs or load shift potentials.

In the course of the decarbonisation of domestic heat demand, it is expected that a large part of the heat will be generated by electricity (e.g. through heat pumps). In order to decarbonise the mobility sector, the aim is to increase the amount of electric vehicles in e.g. Germany from 53,861 in 2018 to 6,000,000 by 2030 \cite{KraftfahrtBundesamt.2020, BMWiBMVBSBMUandBMBF.2011}. Due to the mentioned developments, fundamental characteristics will change in the course of energy demand in the household sector. Furthermore, the introduction of stationary and mobile electricity storage systems as well as stationary heat storage systems enable the storage of energy over periods of single days and therefore open up flexibility potentials in the residential sector, which can support the integration of fluctuating renewable energies. To evaluate these flexibility potentials, data are required that contain information about the mobility behavior of individuals over several days and about their energy relevant \textit{at home} activities. TUD only provide information on activity patterns of two or three individual days, therefore longer-term dependencies in behavior that extend over several days are not captured in existing TUD based models \cite{Torriti.2014}. The objective of this study is to develop an approach which captures long-term dependencies in behavior in order to be able to provide high quality mobility and activity data of individuals to the general public.

Therefore, the paper is structured as follows. Section \ref{sota} provides an overview of the literature in the field of categorical (activity) sequence modelling. Section \ref{method} explains the methodology introduced to capture long-term dependencies in occupancy behavior, before Section \ref{results} presents and discusses the results. The conclusion can be found in Section \ref{conclusion}.

\section{Categorical (activity) sequence modelling}
\label{sota}

Activity schedules are categorical time series in which the state space is defined on the basis of the possible activity states that a person can be in. The most commonly used approaches to model activity sequences is to describe them as Markov chains. \cite{Richardson.2008} developed an approach which uses a first order Markov chain and distinguishes between the states ‘active at home’ and ‘not active at home’ for each person of a household. First order Markov models are adequately suited to describe processes that full fill the Markov property, which refers to the memorylessness of a stochastic process. This means that the transition to a subsequent state depends only on the current state and is independent of previously observed states. It is obvious that residential activity schedules represent more complex processes and therefore cannot easily be represented by a first order Markov model. To overcome this problem, a variety of more complex Markov models have been presented in recent years (semi-Markov \cite{Wilke.2013}, higher order \cite{Flett.2016}, variable memory length \cite{RamirezMendiola.2019}). However, two serious remain with higher-order Markov chains. The number of free parameters in the model increases exponentially with the order of the model and the collection of all possible full high-order Markov chain models is limited and completely stratified. Due to these issues, Markov models are not used in this study. Social practice theory literature points out that in order to understand people’s daily/weekly activity schedules these should be treated as a whole \cite{Shove.2012, Torriti.2017}. As a result, model approaches are required whose memory mechanisms enables complex relationships in human behavior to be recorded as comprehensively as possible.

Attention based neural networks represent the state of the art in the rapidly developing field of natural language processing (NLP) over the past years \cite{Vaswani.12.06.2017, Brown.28.05.2020}. In these networks language is interpreted as categorical sequences in which letters or words represent states. attention based models have shown that they can learn complex long-term dependencies in language and are therefore promising for the application in this study.

\section{Generation of synthetic weekly activity schedules}
\label{method}

\begin{figure}[ht]
\includegraphics[width=1\textwidth]{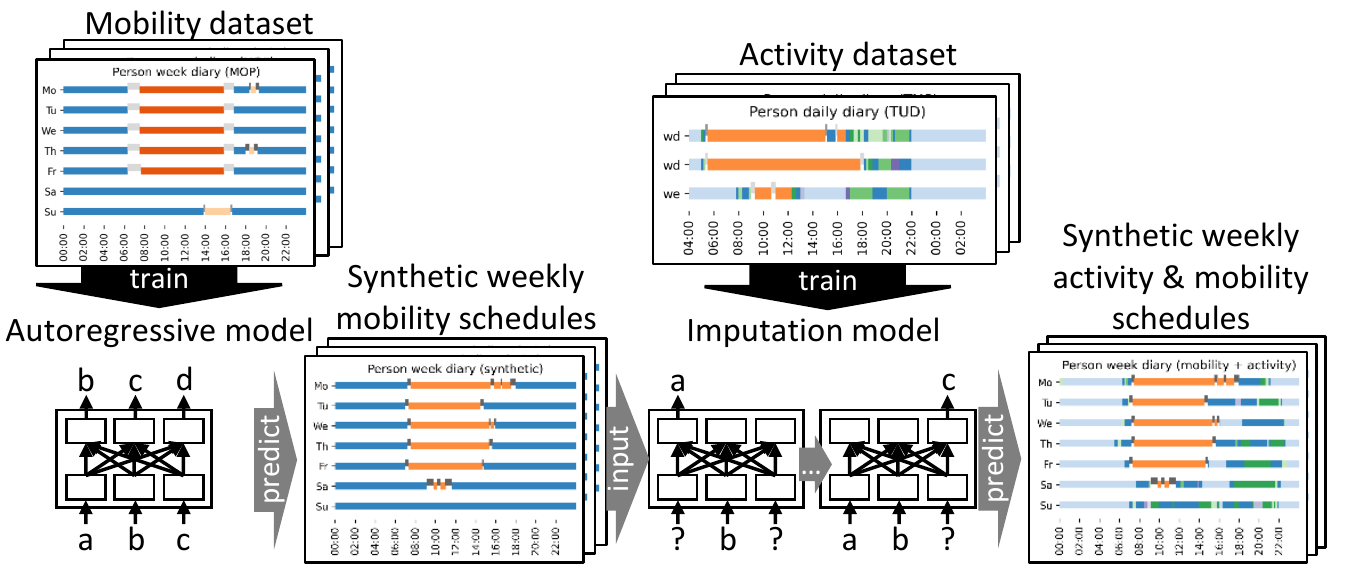}
\caption{Visualization of the two step approach for the generation of weekly activity schedules}
\label{fig.model_overview}
\end{figure}

Figure \ref{fig.model_overview} describes the architecture of the two step approach presented in this study to combine weekly mobility schedules with daily activity schedules for the generation of synthetic weekly activity schedules, which combine the advantages of both datasets. For the application in this study, data from the German Mobility Panel (MOP) and German time use data (TUD) are used. The datasets are described in the appendix in Section \ref{data}.

\begin{figure}[hbt!]
\includegraphics[width=1\textwidth]{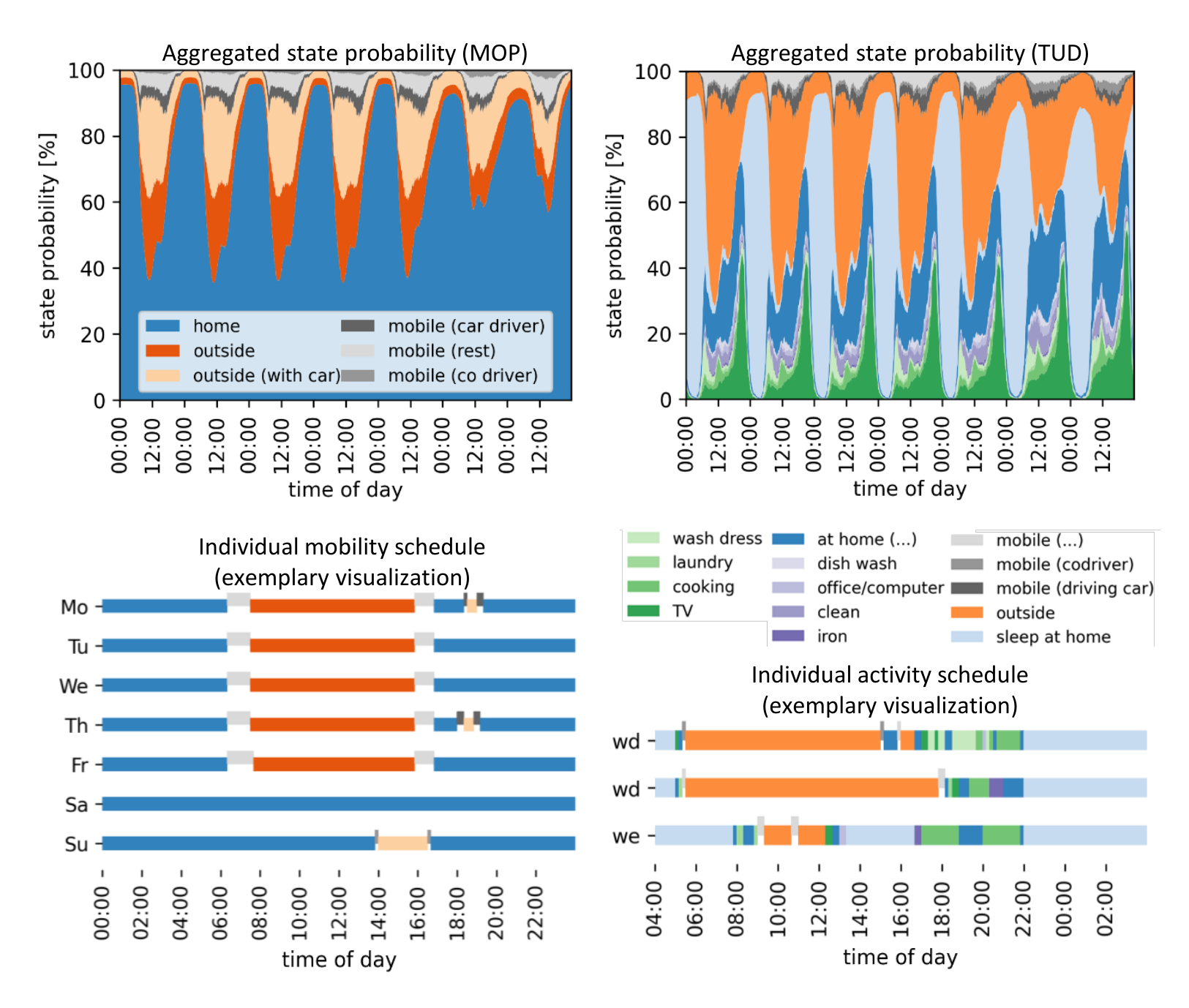}
\caption{Visualization of aggregated state probabilities based on the MOP \cite{Wei.2016} and german TUS \cite{RDCoftheFederalStatisticalOfficeandStatisticalOfficesoftheLander.2002} and exemplary artificial individual diary entries}
\label{fig.input_data}
\end{figure}

\begin{figure}[hbt!]
\includegraphics[width=1\textwidth]{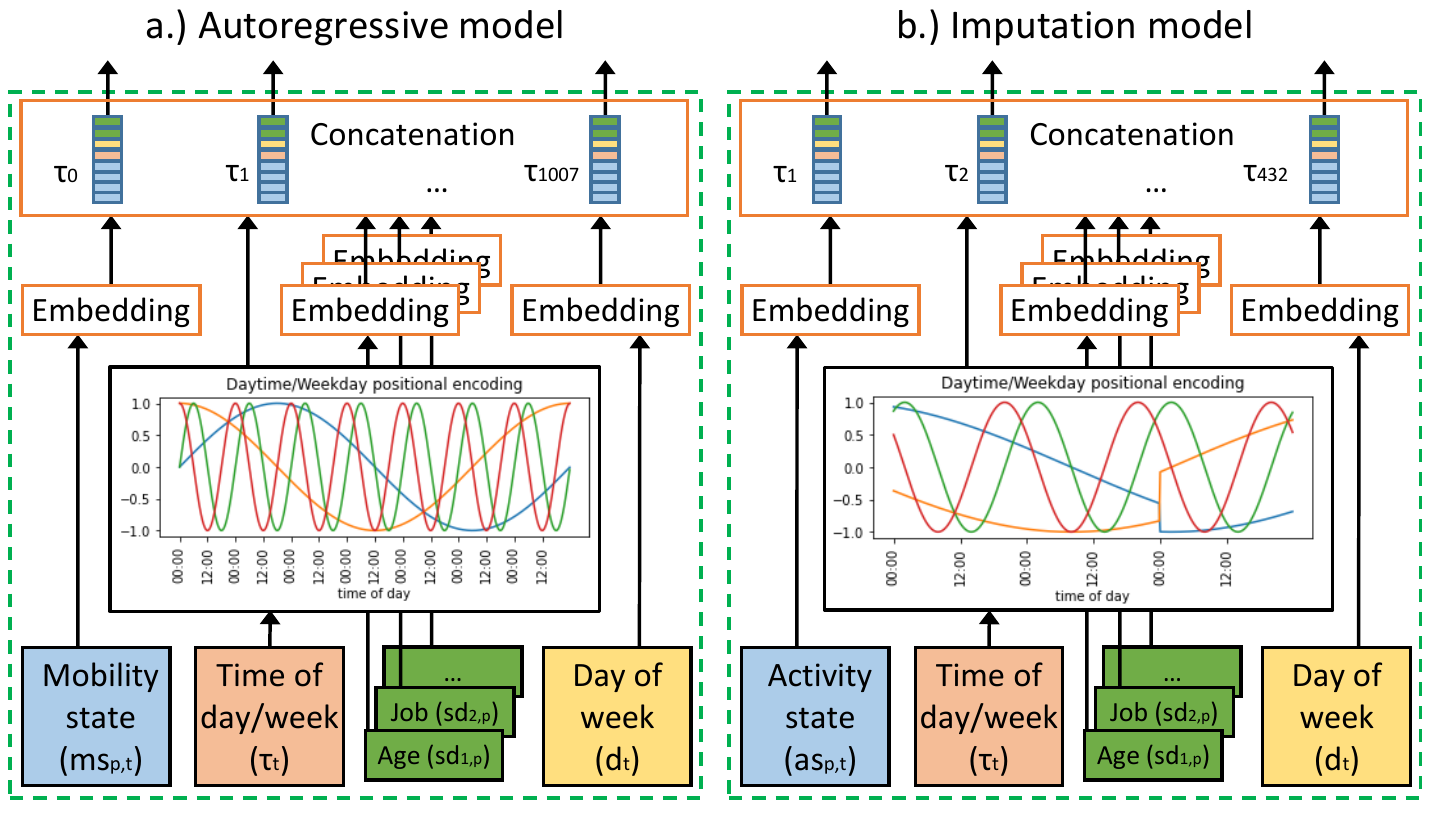}
\caption{Visualization of the training input of the autoregressive model and imputation model and visualization of their first layers}
\label{fig.input_layers}
\end{figure}

\begin{figure}[hbt!]
\includegraphics[width=0.6\textwidth]{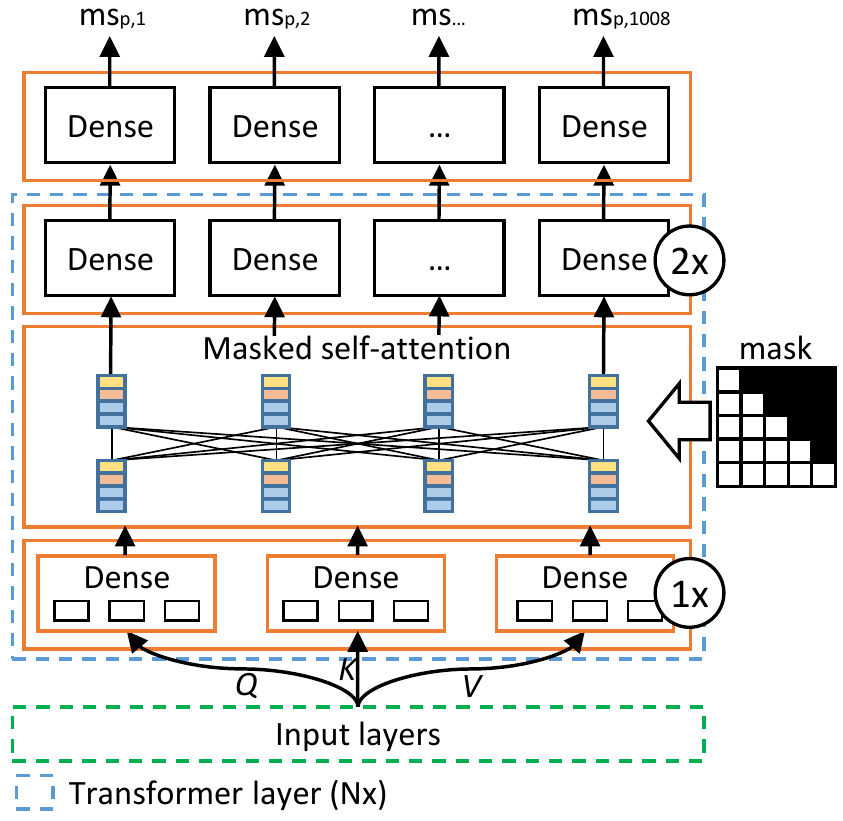}
\centering
\caption{Attention based autoregressive model architecture (residual connections are not visualized)}
\label{fig.autoregressive_model}
\end{figure}

The autoregressive model tries to capture the stochasticity in mobility behavior and to generate, step by step, synthetic mobility schedules that have identical properties as the empirical data. Mobility state information is provided as input in 10-minute resolution, with six mobility states being distinguished from one another (see Figure \ref{fig.input_data}). In contrast to recurrent models (e.g. LSTM models), the temporal relationships in time series must be learned from scratch. To make this easier, timestamp information is provided to the model in the form of sinusoidal position encoding \cite{Vaswani.12.06.2017} as well as weekday embeddings (see Figure \ref{fig.input_layers}). Weekday embeddings are multi-dimensional representations of the weekdays in a continuous space, which are learned during the training process. In this way, daily and weekly rhythms in behavior and differences in behavior on work and weekend days can be learned more easily. Since the socio-demographic composition of both data sets differ and in order to be able to generate socio-demographically differentiated activity schedules, information about the age and occupation type is provided in the form of seven age classes and seven occupation classes. After all time-step-specific information are concatenated, they flow into the main part of the attention based autoregressive model, which is shown in Figure \ref{fig.autoregressive_model}. By using the self-attention mechanism, all dependencies between the 1008 10-minute time steps are learned during the training process. The use of the look-ahead mask ensures that only information from previous time steps is used when predicting the mobility state of the next time step.

The imputation model enriches the “at home” state in the generated mobility schedules by distinguishing between ten different energy relevant \textit{at home} activities (see in Figure \ref{fig.input_data}). During the training process TUD are used which provide activity information about maximum three weekdays (from 4 am to 4 am) for each individual (see Figure \ref{fig.input_layers}). In contrast to the autoregressive model, no look-ahead mask is used in the imputation model, since the model is supposed to use information about future mobility states while predicting current \textit{at home} states. Therefore, only future \textit{at home} states are masked during the training process, so that the attention matrix calculates all dependencies between the unknown \textit{at home} state under consideration and all mobility states as well as all known (previous) energy relevant \textit{at home} states. By using all three days per individual in one sample during the training process, the model is able to learn day-to-day dependencies between \textit{at home} activities (e.g. people go to bed or eat at similar times on consecutive days). During the prediction process, the entire weekly mobility plan is given as input and the \textit{at home} activities are enriched chronologically.

Both models are trained using the cross entropy loss function and the Adam optimizer. The datasets are split up into training data (9-fold cross validation (80 \% training, 10 \% validation)) and test data (10 \%). The generated mobility and activity schedules are constantly evaluated on an individual and aggregated level by calculating the aggregated state probability, the distribution of state durations, the autocorrelation, the amount of weekly activities for each state and the distribution of the hamming distance between all working days over all generated samples.

\section{Results}
\label{results}

\begin{figure}
\includegraphics[width=1\textwidth]{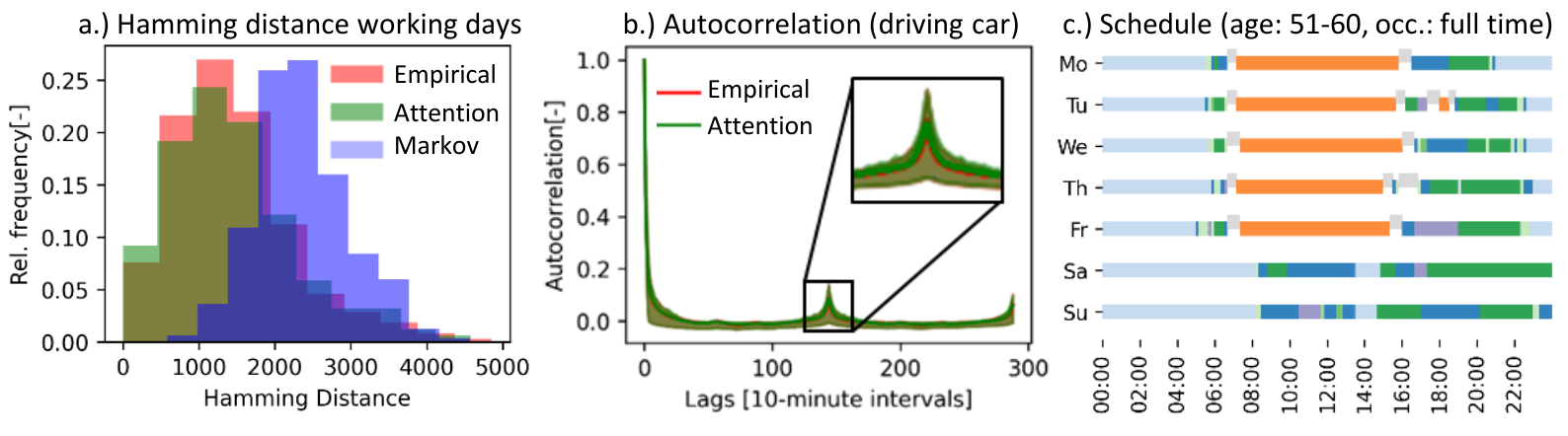}
\caption{Visualization of the distribution of the hamming distance between all working days (a.), the autocorrelation of the state \textit{driving car} (b.) and an example synthetic activity schedule (c.)}
\label{fig.results}
\end{figure}

The model configurations used for the results in Figure \ref{fig.results} can be found in Table \ref{results_autoreg} (model no. 3) and Table \ref{results_impu} (model no. 4). As a reference model for the autoregressive mobility schedule generation, a 1st order Markov model is used. The 1st order Markov model characteristics are representative for the models presented in section \ref{sota}, since marginal changes in the metrics can be achieved by using more complex Markov chains, but the basic problems remain (no long-term memory). From the visualization of the distribution of the hamming distances between the five working days of the week, it can be seen that the attention based models depict similar behavior between the days well, in contrast to the Markov model, in which working days of individuals are not as similar as in the empirical data. In the course of the autocorrelation of the \textit{driving car} state (Figure \ref{fig.results} b.) a peak can be seen after 144 10-minute intervals (24 hours). This peak can be explained by daily rhythms in the commuting behavior of car drivers. From the distribution of the Hamming distance and the 25/75\% quantiles of the autocorrelation, it can be seen that both the diversity between individuals and the average mobility behavior are captured by the model. The described day to day dependencies can also be recognized in the exemplary activity plan in Figure \ref{fig.results} c. The person under consideration leaves home during working days at around the same time and sleeps a bit longer on the weekend. In Tables \ref{results_autoreg} and \ref{results_impu}, error values are given for the various metrics in comparison to the empirical data. In contrast to Markov models, which, based on their structure, meet aggregate state probabilities and the average number of weekly activities well, the attention based models reflect intrapersonal dependencies significantly better, which can be seen by the lower errors in the duration of the states, the autocorrelation and the Hamming distance. The errors of the attention-based model in the aggregated state probabilities are, as can be seen from Table \ref{results_autoreg} and Table \ref{results_impu}, slightly higher than those of the Markov model, in which the error tends to zero with increasing sample size. However, the average error in the aggregated state probability over all socio-demographic groups between the generated data and the empirical mobility data is more than two times lower than the error that arises when comparing the overlapping states of the two input data sets. It can be concluded that the approach presented enables to combine the advantages of weekly mobility data with a large sample size with the advantages of high-resolution activity data of individual days, thus creating a new data basis which can be used for further analyses of human occupancy and mobility behavior. 

\section{Conclusion}
\label{conclusion}

Over the past years, more and more models have been published that aim to capture relationships in human residential behavior. Most of these models are different Markov variants or regression models that have a strong assumption bias and are therefore unable to capture complex long-term dependencies and the diversity in occupant behavior. This work shows that attention based models are able to capture complex long-term dependencies in occupancy behavior and at the same time adequately depict the diversity in behavior across the entire population and different socio-demographic groups. By combining an autoregressive generative model with an imputation model, the advantages of two data sets are combined and new data are generated which are beneficial for multiple use cases (e.g. generation of consistent household energy demand profiles). The two step approach generates synthetic activity schedules that have similar statistical properties as the empirical collected schedules and do not contain direct information about single individuals. Therefore, the presented approach forms the basis to make data on occupant behavior freely available, so that further investigations based on the synthetic data can be carried out without a large data application effort. In future work it is planned to take interpersonal dependencies into account in order to be able to generate entire household behavior profiles.

\begin{ack}
This work was supported by the Helmholtz Association under the Joint Initiative “Energy Systems Integration” (funding reference: ZT-0002) and was done during a research stay funded by the Centre for Research into Energy Demand Solutions (CREDS) at the University of Reading (UK). This work was supported by UKRI [grant numbers EP/R000735/1, EP/R035288/1 and EP/P000630/1].
\end{ack}

\bibliographystyle{plain}
\bibliography{references}

\section*{Broader impact on climate change}

In 2017, the residential sector accounted for 27\% of European final energy consumption (mobility excluded) and therefore takes a key role in achieving European climate targets \cite{Eurostat.2019}. Approximately 64\% of the residential final energy demand can be attributed to space heating demand, 15\% to domestic hot water demand and 20\% to the demand for lighting, cooking and appliances \cite{Eurostat.2019}. In the course of the decarbonisation of domestic heat demand, it is expected that a large part of the heat will be generated by electricity (e.g. through heat pumps). In order to decarbonise the mobility sector, the aim is to increase the amount of electric vehicles in Germany from 53,861 in 2018 to 6,000,000 by 2030 \cite{KraftfahrtBundesamt.2020, BMWiBMVBSBMUandBMBF.2011}. Due to the mentioned developments and an expected further increase in photovoltaic battery systems in the residential sector, fundamental characteristics will change in the course of energy demand in the household sector. Therefore, residential neighbourhoods are gaining increased attention by policy makers. In order to establish targeted policy interventions for different households, which allow an optimal integration of low carbon technologies and open up flexibility options on the demand side, fundamental factors influencing the structure of energy demand must be understood.   
The diversity and differences in electricity and heat consumption between different households/buildings is a major obstacle to understanding future energy consumption.

\section*{Broader impact regarding ethical aspects}

When providing behavioral data, ethical aspects such as data privacy must be taken into account at all times. The data sets used in this work do not allow any conclusions to be drawn back to individuals. However, before the synthetic data are made available online, differential privacy of the models must be ensured.

The data set provided by this work can support energy system planners, for example in predicting future load peaks due to high electrical demand at certain times. Furthermore, politicians can be supported in the introduction of variable electricity tariffs so that they do not favor or disadvantage different socio-demographic groups.Based on the behavior of different socio-demographic groups, the data could be used in the marketing sector to target specific groups that increasingly watch television at certain times. When using synthetic data, it must also be taken into account that models always lead to errors and that the synthetic data therefore differ from the empirically collected data, which are also subject to certain biases. This study particularly points out that the two basic data sets differ relatively strongly in terms of the composition of the socio-demographic groups. This is taken into account when generating the synthetic data by using socio-demographic factors age and occupation status while merging the datasets.

\section{Appendix}

\subsection{Data}
\label{data}
Information about the weekly mobility behavior is taken from the German Mobility Panel (MOP) which collects information about each travel activity of about 1,500 to 3,100 individuals since 1994 every year \cite{Wei.2016, Zumkeller.2009}. In this study 26,610 weekly mobility schedules from the years 2001 till 2017 together with their associated socio-demographic information (age, occupation) are used as input. Information about energy relevant \textit{at home} activities is taken from the Harmonized European Time Use Survey \cite{RDCoftheFederalStatisticalOfficeandStatisticalOfficesoftheLander.2002, Eurostat.2000}. Activity diaries and socio-demographic information of 11,921 individuals out of 5,443 households are used to train the imputation model. Most of the participants provide diaries on two weekdays and one weekend-day in 10-minute resolution. Figure \ref{fig.input_data} shows the temporal course of the aggregated state probability over the entire dataset population and an exemplary mobility and activity schedule.

\subsection{Figures and tables}

\begin{figure}[hbt!]
\includegraphics[width=0.6\textwidth]{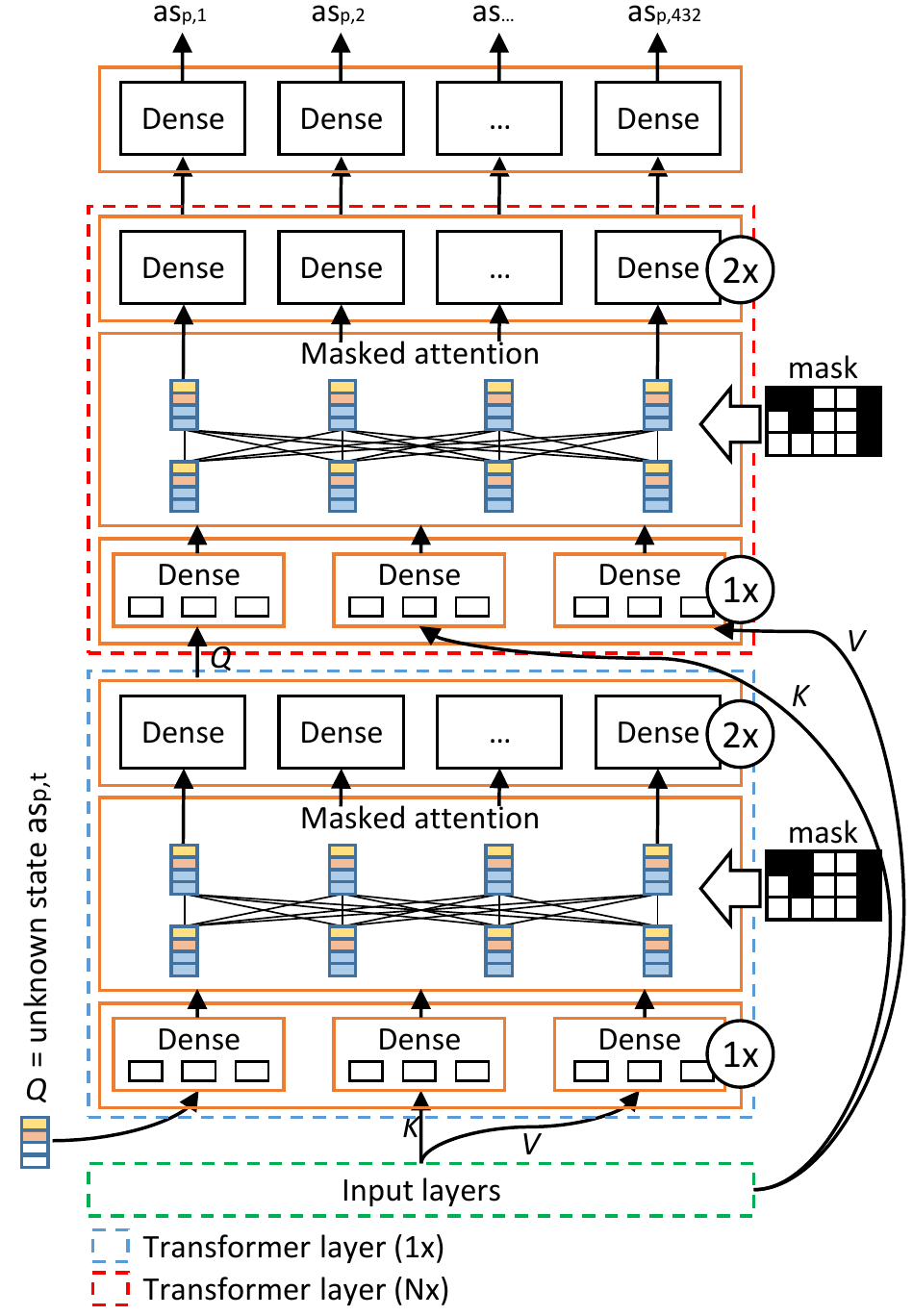}
\centering
\caption{Attention based imputation model architecture (residual connections are not visualized)}
\label{fig.imputation_model}
\end{figure}

\begin{figure}[hbt!]
\includegraphics[width=1\textwidth]{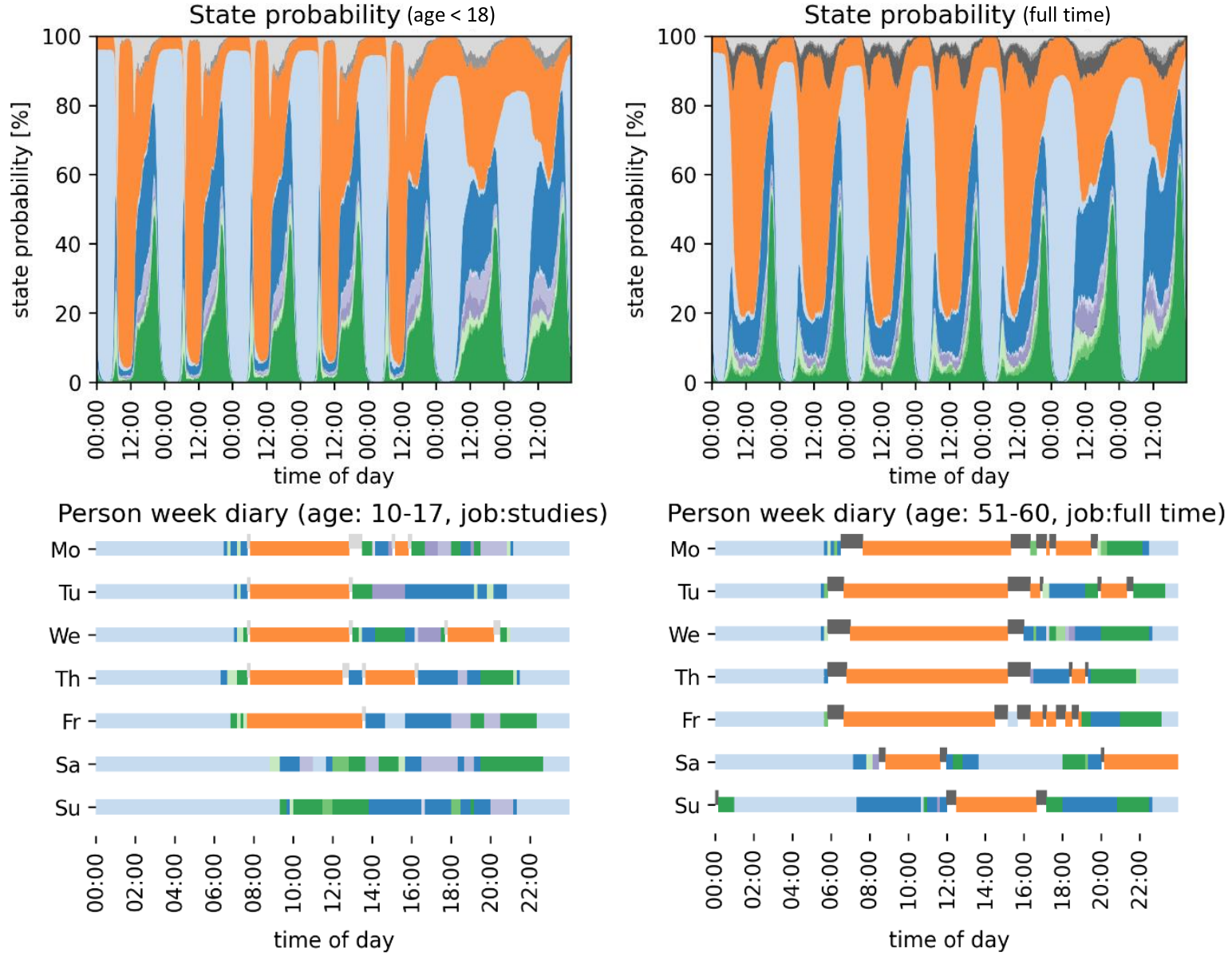}
\centering
\caption{The top two figures represent the course of the aggregated state probability for 1,500  generated activity schedules for persons under 18 years of age and for full time employees. The lower two representations are two exemplary activity schedules for a person under the age of 18 and a full-time employee (a legend can be found in Figure \ref{fig.input_data})}
\label{fig.results_appendix}
\end{figure}

\begin{table}[p]
  \caption{Hyperparameter configurations and metrics for the attention based autoregressive model. Metrics are calculated using N = 2,000 samples.}
  \label{results_autoreg}
  \centering
  \begin{tabular}{llllllllll}
    \toprule
    No.     & layers/   & sp & sd & ac & na & hd & loss & acc. & epoch\\
            & d\_model/& rmse & rmse & rmse & mae & mae    \\
            & learning rate/ & [\%] & [\%] & [-] & [-] & [-] & [-] & [\%] & [-]   \\
            & batch size/ \\
    \midrule
    1     & 1/64/0.001/64  & 0.83	&0.31	&1.32	&2.96	&244	&0.14	&95.95	&9\\
    2     & 4/64/0.001/64  & 0.91	&0.16	&0.70	&2.53	&33	    &0.128	&96.34	&15\\
    3     & 8/64/0.001/64  & 0.86	&0.17	&0.54	&3.6	&5	    &0.127	&96.36	&7\\
    4     & 4/128/0.001/128& 0.89	&0.24	&0.59	&3.60	&9	    &0.128	&96.33	&6\\
        & 1st order Markov & 0.53  & 0.53 & 3.79 & 0.73 & 908 \\
    \bottomrule
  \end{tabular}
\end{table}

\begin{table}[p]
  \caption{Hyperparameter configurations and metrics for the attention based imputation model. Metrics are calculated using N = 2,000 samples.}
  \label{results_impu}
  \centering
  \begin{tabular}{lllllllll}
    \toprule
    No.     & layers/   & sp & sd & ac & na  & loss & acc. & epoch\\
            & d\_model/& rmse & rmse & rmse & mae     \\
            & learning rate/ & [\%] & [\%] & [-]  & [-] & [-] & [\%] & [-]   \\
            & batch size/ \\
    \midrule
    1	&1/64/0.001/256	&0.58	&0.39	&0.50	&0.50	&0.469	&86.97	&158\\
    2	&4/64/0.001/256	&0.58	&0.39	&0.44	&0.62	&0.436	&87.32	&22\\
    3	&4/64/0.001/64	&0.57	&0.38	&0.36	&0.90	&0.436	&87.35	&8\\
    4	&4/64/0.0005/128&0.49	&0.39	&0.39	&0.66	&0.431	&87.41	&37\\

    \bottomrule
  \end{tabular}
\end{table}

\end{document}